\documentclass[onecolumn]{aastex631}

\usepackage{CJK}

\usepackage{xcolor}

\shorttitle{repeating frb$\rm s $ from tidal disruption}
\shortauthors{Abdusattar Kurban et al.}
\begin{document}
	
\begin{CJK*}{UTF8}{gbsn}
		
\title{Periodic repeating fast radio bursts: interaction between a magnetized neutron star and its planet in an eccentric orbit}
		
		
\correspondingauthor{Abdusattar Kurban, Yong-Feng Huang}
\email{akurban@xao.ac.cn, hyf@nju.edu.cn}
		
\author[0000-0002-2162-0378]{Abdusattar Kurban (阿布都沙塔尔·库尔班)}
		
\affil{Xinjiang Astronomical Observatory, Chinese Academy of Sciences, Urumqi 830011, Xinjiang, People's Republic of China}
\affil{School of Astronomy and Space Science, Nanjing University, Nanjing 210023, People's Republic of China}
\affil{Key Laboratory of Radio Astronomy, Chinese Academy of Sciences, Urumqi 830011, Xinjiang, People's Republic of China}
\affil{Xinjiang Key Laboratory of Radio Astrophysics, Urumqi 830011, Xinjiang, People's Republic of China}

\author[0000-0001-7199-2906]{Yong-Feng Huang (黄永锋)}
\affil{School of Astronomy and Space Science, Nanjing University, Nanjing 210023, People's Republic of China}
		
\affiliation{Key Laboratory of Modern Astronomy and Astrophysics (Nanjing University),
Ministry of Education, Nanjing 210023, People's Republic of China}
		
\author[0000-0001-9648-7295]{Jin-Jun Geng (耿金军)}
\affil{Purple Mountain Observatory, Chinese Academy of Sciences, Nanjing 210023, People's Republic of China}
		
\author[0000-0002-0238-834X]{Bing Li (李兵)}
\affil{Key Laboratory of Particle Astrophysics, Chinese Academy of Sciences, Beijing 100049, People's Republic of China}
\affil{Particle Astrophysics Division, Institute of High Energy Physics, Chinese Academy of Sciences, Beijing 100049, People's Republic of China}
		
\author[0000-0001-7943-4685]{Fan Xu (许帆)}
\affil{School of Astronomy and Space Science, Nanjing University, Nanjing 210023, People's Republic of China}
		
\author[0000-0002-1793-047X]{Xu Wang (王旭)}
\affil{School of Astronomy and Space Science, Nanjing University, Nanjing 210023, People's Republic of China}

\author[0000-0003-4686-5977]{Xia Zhou (周霞)}
\affil{Xinjiang Astronomical Observatory, Chinese Academy of Sciences, Urumqi 830011, Xinjiang, People's Republic of China}
\affil{Key Laboratory of Radio Astronomy, Chinese Academy of Sciences, Urumqi 830011, Xinjiang, People's Republic of China}
\affil{Xinjiang Key Laboratory of Radio Astrophysics, Urumqi 830011, Xinjiang, People's Republic of China}

\author[0000-0003-1845-4900]{Ali Esamdin (艾力·伊沙木丁)}
\affil{Xinjiang Astronomical Observatory, Chinese Academy of Sciences, Urumqi 830011, Xinjiang, People's Republic of China}

\author[0000-0002-9786-8548]{Na Wang (王娜)}
\affil{Xinjiang Astronomical Observatory, Chinese Academy of Sciences, Urumqi 830011, Xinjiang, People's Republic of China}
\affil{Key Laboratory of Radio Astronomy, Chinese Academy of Sciences, Urumqi 830011, Xinjiang, People's Republic of China}
\affil{Xinjiang Key Laboratory of Radio Astrophysics, Urumqi 830011, Xinjiang, People's Republic of China}





\begin{abstract}
	
	Fast radio bursts (FRBs) are mysterious transient phenomena. The study of repeating
	FRBs may provide useful information about their nature due to their redetectability.
	The two most famous repeating sources are FRBs 121102 and 180916, with a period of
	157 days and 16.35 days, respectively. Previous studies suggest that the
	periodicity of FRBs is likely associated with neutron star (NS) binary systems.
	Here we introduce a new model which proposes that periodic repeating FRBs are due to the interaction
	of a NS with its planet in a highly elliptical orbit. The periastron of the planet is
	very close to the NS so that it would be partially disrupted by tidal force every time
	it passes through the periastron.
	Fragments generated in the process could interact with the compact star
	through the Alfv\'{e}n wing mechanism and produce FRBs.
	The model can naturally explain the repeatability of FRBs with a period ranging
	from a few days to several hundred days, but it generally requires that the
	eccentricity of the planet's orbit should be large enough. Taking FRBs 121102 and
	180916 as examples, it is shown that the main features of the observed repeating
	behaviors can be satisfactorily accounted for.
	
\end{abstract}






\section{Introduction} \label{sec:intro}


The first discovery of fast radio bursts (FRBs) by \citet{Lorimer2007Sci} and
consequent reports of five similar sources by \citet{Keane2012MNRAS} and
\citet{Thornton2013Sci} opened a new window in astronomy.
Since then FRBs have become an active topic for research.
The isotropic energy released by FRBs is in the range of $ 10^{38} - 10^{46} $
erg, and their duration is typically several milliseconds. The
observed dispersion measure is $\rm  \sim 110 - 2596 \,pc \,cm^{-3} $
\citep{Petroff2019AARv}, which strongly hints that FRBs are of cosmological origin.
According to the observed repeatability \citep{Petroff2015MNRAS}, these enigmatic events may come from
two kinds of progenitors, i.e. repeating sources and nonrepeating sources.

Many models (see \citet{Platts2019PhR} for a recent review) have
been proposed to interpret the properties of FRBs.
However, their underlying physics -- the progenitor as well as emission mechanism -- remains unclear
\citep{Katz2018PrPNP,Petroff2019AARv,Platts2019PhR,Cordes2019ARAA,Zhang2020Natur}.
Repeating FRBs, in particular periodic repeating FRBs, may provide valuable information
about the nature of this mysterious phenomenon.

Here we will mainly focus on the periodic repeating activities of FRBs.
The most famous periodic repeating sources are FRB 121102
and FRB 180916.
FRB 121102 has a period of 157 days \citep{Rajwade2020MNRAS},
and FRB 180916 has a period of 16.35 days \citep{Chime/Frb2020Natur}.
Two kinds of models, the single-star model and binary model, have been proposed to
interpret the periodic repeatability of these FRBs. The single-star models are
mainly concerned with the precession of neutron
stars (NS) \citep{Levin2020ApJ,Yang2020ApJ,Sob'yanin2020MNRAS,Zanazzi2020ApJ} while
the binary models associate FRBs with the interaction between the two objects in
NS binary systems
\citep{Mottez2014AA, Dai2016, Zhang2017ApJ, Zhang2018ApJ, Lyutikov2020ApJ, Ioka2020ApJ, Dai2020a, Dai2020,
	Gu2020MNRAS, Mottez2020AA, Geng2020, Decoene2021AA, Du2021MNRAS}.
Usually, the precession period of NS is unlikely to be as long as 16.35 days~\citep{Chime/Frb2020Natur}.
Additionally, the fixed emission region of FRBs in
the precession models has not yet been properly addressed \citep{Xiao2021SCPMA} .
Various observational facts imply that binary models are more likely favored by
the periodicity of FRBs. The binary-interaction models can be further categorized into two
main classes: wind-like models and accretion/collision-like models.
The wind-like models include the binary comb mechanism \citep{Zhang2017ApJ,Zhang2018ApJ,Ioka2020ApJ},
mild pulsars in tight O/B-star binaries \citep{Lyutikov2020ApJ},
small bodies orbiting around a pulsar or a magnetar \citep{Mottez2014AA, Mottez2020AA, Voisin2021MNRAS},
and Kozai-Lidov feeding of NSs in binary systems \citep{Decoene2021AA}.
The collision/accretion-like models include the
collision between a magnetized NS and an asteroid belt \citep{Dai2016, Smallwood2019, Dai2020a, Dai2020},
accretion of strange stars from low-mass companion stars \citep{Geng2021}, and NS-white dwarf (WD) interactions
\citep{Gu2016ApJ, Gu2020MNRAS}. FRBs and their counterparts in other wavelengths have been studied by
\citet{Yang2021ApJa},\citet{Yang2021ApJb}, and by many other authors.
As suggested earlier by a few authors, collisions between small bodies and a NS can generate transient
events such as gamma-ray bursts \citep{Campana2011Natur}, glitch/anti-glitches and X-ray
bursts \citep{Huang2014, Yu2016RAA}, and FRBs \citep{Geng2015,Dai2016}.

Tidal disruption of minor planets/asteroids around WDs has also been
extensively studied \citep{Bear2013NewA, Vanderburg2015Natur, Granvik2016Natur}.
Recent simulations \citep{Malamud2020a,Malamud2020b} have showen that a planet in a highly
eccentric orbit around a WD could be tidally disrupted by tidal force, and materials
in the inner side of the orbit would be accreted by the WD. Accreted clumps of such
materials may be responsible for the pollution of a WD's atmosphere by heavy
elements \citep{Vanderburg2015Natur, Malamud2020a, Malamud2020b}.
Similar processes (disruption of a planet) can also
occur in NS-planet systems if the initial parameters of the planetary system fulfill
the tidal disruption condition \citep{Brook2014ApJ}. In fact, GRB 101225A may occur in this way \citep{Campana2011Natur}.
Much efforts have also been made to search for close-in exoplanets around
pulsars \citep{Geng2015, Huang2017ApJ, Kuerban2020ApJ}.

In this study, we propose a new model to explain the periodic repeating properties of FRB
sources. We argue that when a planet is in a highly eccentric orbit around a NS,
it would be partially disrupted every time it passes through the pericenter.
The major fragments generated during the disruption will interact with the pulsar
(rotating NS) wind to produce a series of FRBs. This model can naturally explain
the periodic behavior of repeating FRBs. The structure of our paper is as follows.
In Section \ref{sec:model}, we present the basic framework of our model for repeating FRBs.
In Section \ref{sec:wind-clump interaction}, the wind-clump interaction mechanism for FRBs is introduced.
In Section \ref{sec:periodicity-active}, the periodicity and active window are described
in view of the model.
In Section \ref{sec:evaporation}, we estimate the evaporation timescale for a planet in an elliptical orbit.
In Section \ref{sec:formation of eccentric orbit}, we address the possible existence
of pulsar planets in highly eccentric orbits.
Finally, Section \ref{sec:conclusion} presents our conclusions and some brief discussion.

\section{Model}\label{sec:model}

The planet-disruption interpretation for the pollution of a WD's atmosphere by heavy
elements \citep{Vanderburg2015Natur,Granvik2016Natur,Stephan2017ApJ,Malamud2020a,Malamud2020b} and
the Alfv\'{e}n wing theory for FRBs \citep{Mottez2014AA,Mottez2020AA} motivates us to
investigate the periodic repeating
activities of FRBs in the framework of a NS-planet interaction model.
When a planet is in an highly elliptical orbit with the periastron distance being
small enough, it might be partially disrupted every time it passes through the pericenter.
The disrupted fragments formed during this process will regularly interact
with the host NS and produce periodic repeating FRBs.

\begin{figure}
	\plotone{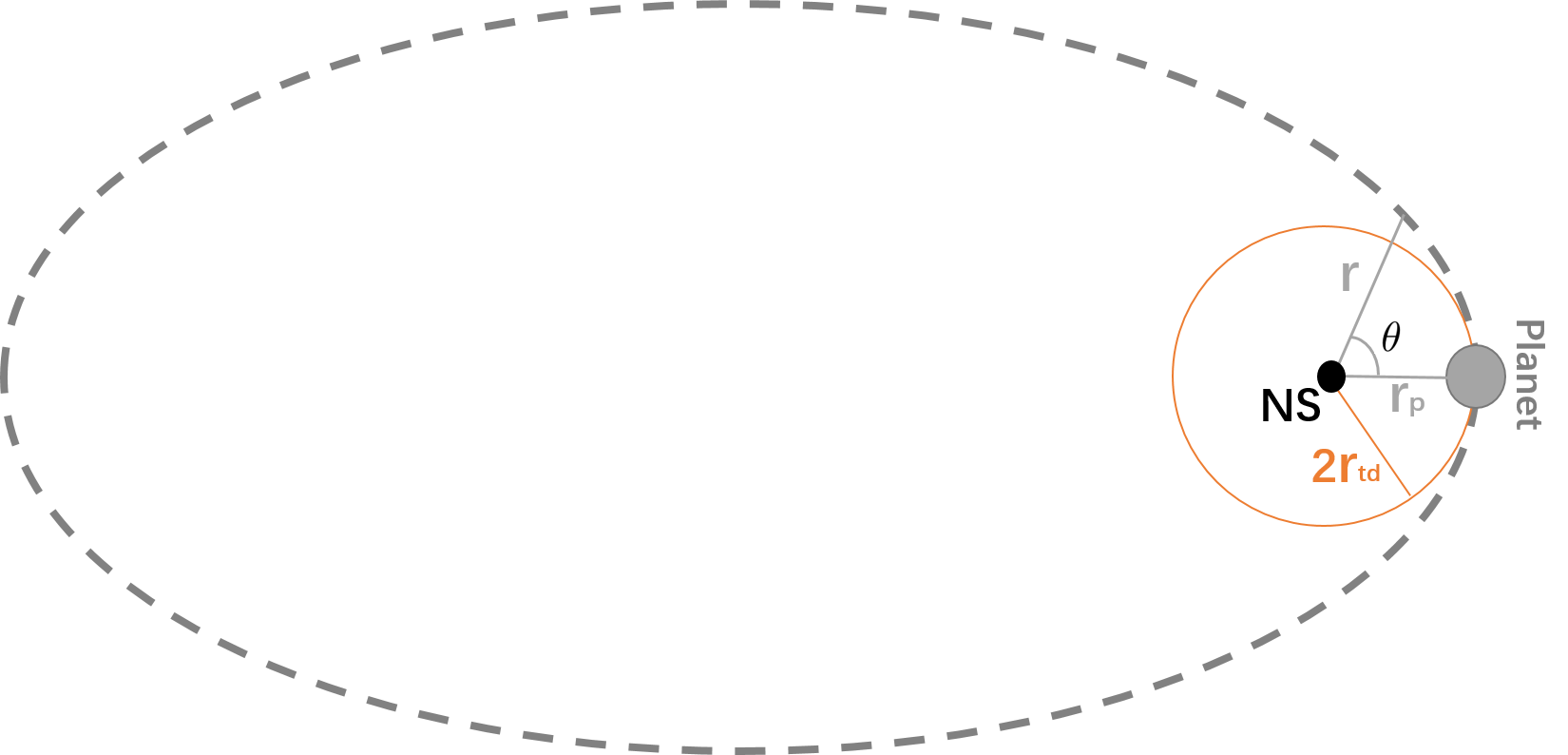}
	\caption{Schematic illustration (not to scale) of a pulsar planet in a highly eccentric
		orbit. The central star is a NS and the planet is assumed to be a typical
		rocky object. $ r $ is the separation between the NS and the planet at phase $ \theta $.
		$ r_{\rm p} $ is the periastron distance of the orbit. $ 2\,r_{\rm td} $ is
		a characteristic distance at which partial tidal disruption will occur
		(see text for more details). }\label{fig:fig1}
\end{figure}

Figure~\ref{fig:fig1} illustrates the general picture of a NS-planet system in an eccentric orbit.
We assume that the central star is a NS with a mass $ M_{\star} = 1.4\,M_{\odot}$, and the companion is a
rocky planet with a mass $ m $, mean density $ \bar\rho $, and an orbital period $ P_{\rm orb} $.
The semi-major axis ($ a $) and orbital period are related by the Kepler's third law as
\begin{eqnarray} \label{eq:kepler_3}
	\frac{P_{\rm orb}^2}{a^3} = \frac{4 \pi^2}{G(M_{\star} + m)}.
\end{eqnarray}
The distance between the NS and planet at phase $ \theta $ (the true anomaly; see Figure~\ref{fig:fig1})
in the eccentric orbit is
\begin{eqnarray} \label{eq:r_orb}
	r = \frac{a(1-e^2)}{1 + e \,\cos\theta},
\end{eqnarray}
where $ e $ is the eccentricity of the orbit.
The characteristic tidal disruption radius of the planet depends on its density as \citep{Hills1975Natur}
\begin{eqnarray} \label{eq:r_td}
	r_{\rm td} \approx \left( \frac{6M_{\star}}{\pi\,\bar \rho}\right) ^{1/3}.
\end{eqnarray}

\begin{figure}
	\plotone{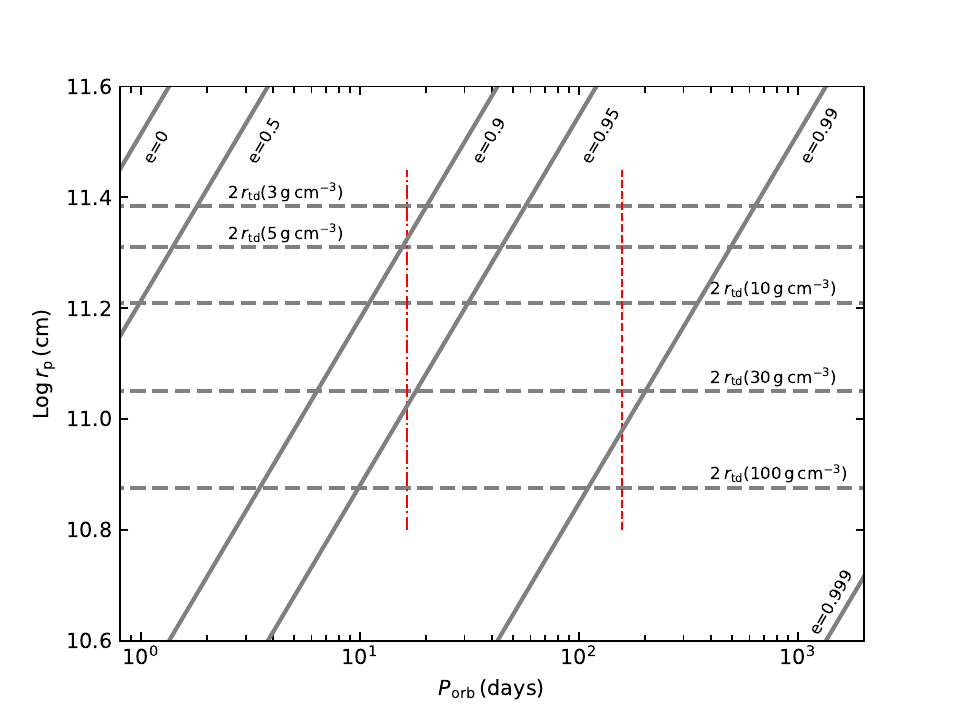}
	\caption{The periastron distance as a function of the orbital period for various
		eccentricities.
		Horizontal lines correspond to $ r_{\rm p} = 2 r_{\rm td} $ for different
		planet densities (marked in the brackets).
		The dashed-dotted vertical line corresponds to an orbital period of 16.35 days,
		and the dashed vertical line is 157 days.} \label{fig:fig2}
\end{figure}

Whether a planet will be tidally disrupted or not depends on its separation ($ r $) with respect to
the NS. If $ r $ is smaller than a critical value of $ 2.7\,r_{\rm td} $,
then it will begin to be partially disrupted \citep{Liu2013ApJ}.
The separation between the planet and NS is different when the planet is at different
orbital phase. At periastron, it is
\begin{eqnarray} \label{eq:r_p}
	r_{\rm p} = a(1 - e).
\end{eqnarray}
For a highly elliptical orbit on which the separation varies in a very wide range,
the planet may be tidally affected mainly near the periastron and it is relatively
safe at other orbital phases. Here, we focus on the disruption near the periastron.
If the orbit is too compact (for example, $ r_{\rm p}\leq r_{\rm td} $), then the
disruption is violent and the planet will be completely destroyed.
However, when $ r_{\rm td} < r_{\rm p} < 2.7\,r_{\rm td} $, then the planet will only
be partially disrupted every time it passes by the periastron.
Since the density at the surface is relatively small, the outer crust of the planet
will be destroyed first, which gives birth to a number of fragments with the size of
a few kilometers. The main portion of the planet will retain its integrity. The idea
of partial disruption has been supported both from observations \citep{Manser2019Sci} and
simulations \citep{Liu2013ApJ, Malamud2020a, Malamud2020b}.

In our study, we assume $ r_{\rm p} = 2\,r_{\rm td} $ for simplicity, which satisfies the
condition for a partial disruption. We can then calculate the relation between the
periastron distance and the orbital period, which depends on the orbital eccentricity.
The results are shown in Figure~\ref{fig:fig2}.
For comparison, we have also marked the partial tidal disruption distance
($2\,r_{\rm td}$ for the planet with a particular density) as horizontal lines.
We can see that a partial disruption would occur near the periastron for a wide range of orbital periods.
For example, for an orbit with $e = 0.95$, the partially disrupted planet will have
an orbital period of $\sim 20$ days when its mean density is 30 g cm$^{-3}$.
If the mean density is 10 g cm$^{-3}$, 5 g cm$^{-3}$, or 3 g cm$^{-3}$, then the
orbital period will be 30 days, 43 days, and 60 days, correspondingly.
More generally, for a planet with a mean density ranging from
$ 3 $ to $ 10 \rm \,g\, cm^{-3} $, partial disruption will
occur for $P_{\rm orb} \sim $ 2 --- 600 days when the eccentricity is $e \sim$ 0.5 --- 0.99.
Note that the mass of the planet does not affect the disruption too much.
The disruption process is mainly determined by orbital parameters and
the mean density of the planet.

A number of fragments will be generated during the partial disruption process.
These fragments will experience some complicated dynamical interactions such
as gravitational perturbation (\citet{Naoz2016ARAA}, see below) and
scattering/collision \citep{Cordes2008ApJ}.
These interactions lead the clumps to orbit around the central NS with slightly different
orbital parameters (velocity, semi-major axis, eccentricity, inclination
relative to the planet's orbit, etc.). In the orbiting process, the interaction between
the clumps and the pulsar wind can generate FRBs through the Alfv\'{e}n wing 
mechanism \citep{Mottez2014AA,Mottez2020AA}.

The above process of partial disruption happens periodically every time the surviving main
portion of the planet passes through the periastron.
Consequently, this regular interaction can account for
the periodic repeating FRBs.

\section{Wind-Clump interaction mechanism}\label{sec:wind-clump interaction}

Orbiting small bodies immersed in the relativistic wind of a highly magnetized
pulsar can be the sources of repeating FRBs \citep{Mottez2014AA,Mottez2020AA}.
The interaction between the small body and the pulsar wind produces a small
Alfv\'{e}n wing angle (see Table 1 in \citet{Mottez2014AA}). When the wind plasma crosses the
Alfv\'{e}n wing, it sees a rotation of the ambient magnetic field that can cause
radio wave instabilities.
In the observer's reference frame, the radiation is focused into a very small solid angle due to the
relativistic beaming effect, which amplifies the flux density and produces FRBs.
At a distance of $\sim 1$ AU from the pulsar, the required size for a small body to produce FRBs is
a few kilometers.

When a pulsar with a surface magnetic field $ B_{\star} $ and angular frequency $ \Omega_{\star} $
interacts with a small body of size $ R_{\rm c} $, it generates an Alfv\'{e}n wing with a power of \citep{Mottez2020A}
\begin{equation}\label{wing power}
\dot{E}_{\rm A} = \frac{\pi}{\mu_{0} c^3} B_{\star}^2 R_{\star}^6 \Omega_{\star}^4 R_{c}^2 r^{-2},
\end{equation}
where $\mu_{0}$ is the magnetic conductivity, $ c $ is the speed of light,
$ R_{\star} $ is the radius of the pulsar,
$ r $ is the separation between the two objects and is a function of $ \theta $ as shown in Eq.~(\ref{eq:r_orb}).
The radio emission power of the Alfv\'{e}n wing is
\begin{equation}\label{wing radio emission}
\dot{E}_{\rm radio} = \epsilon_{\rm r} \dot{E}_{\rm A},
\end{equation}
where $ 2\times 10^{-3} \leq \epsilon_{\rm r} \leq 10^{-2}$ is the radiation efficiency \citep{Zarka2001,Zarka2007}.
In the observer's reference frame, the radio flux density generated from the interaction between
the pulsar wind and a small sized object is \citep{Mottez2020AA}
\begin{equation}\label{wind->frb}
	\left( \frac{S}{\rm Jy}\right) = 2.7 \times 10^{-9} A_{\rm cone} \left( \frac{\gamma}{10^5}\right)^2 \left( \frac{\epsilon_{\rm r}}{10^{-3}}\right) \left( \frac{R_{\rm c}}{\rm 10^9 cm}\right)^2 \left( \frac{r}{\rm AU}\right)^{-2} \left( \frac{R_{\star}}{\rm 10^6 cm}\right)^6 \left( \frac{B_{\star}}{\rm 10^9 G}\right)^2 \left( \frac{P_{\star}}{\rm 10^{-2} s}\right)^{-4}
	\left( \frac{D}{\rm 1 Gpc}\right)^{-2} \left( \frac{\Delta \nu}{\rm 1 GHz}\right)^{-1},
\end{equation}
where $ \gamma $ is the Lorentz factor of the pulsar wind,
$ \Delta \nu $ is the emission bandwidth,
$ D $ is the luminosity distance,
$ P_{\star} $ is the spin period of the pulsar.
In Eq.~(\ref{wind->frb}), $ A_{\rm cone} = 4\pi/\Omega_{\rm A}$ is an indication of the beaming factor.
The radio waves are emitted into a solid angle $ \Omega_{\rm A}$ in the source frame, which could be
nearly isotropic. 
Note, however, that the radiation is limited in
a solid angle $ (\Omega_{A}/4) \gamma^{-2} $ in the observer's reference frame due to the beaming effect; 
the observer can see the FRBs only when the radio beams point toward the observer.

\begin{figure}[!h]
	\plotone{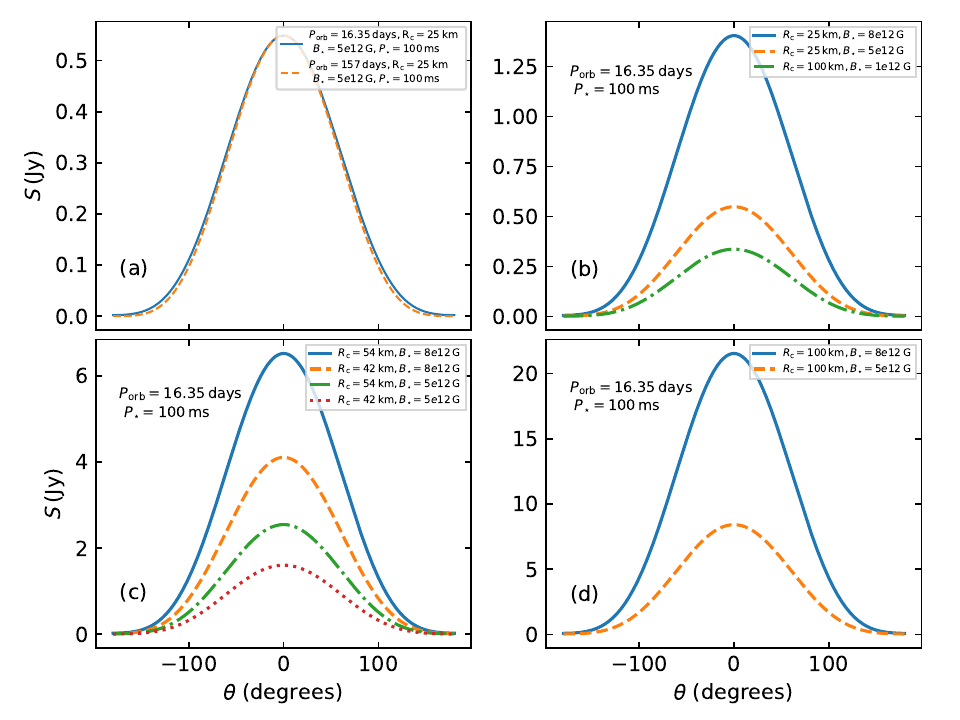}
	\caption{The flux density of FRBs vs. the orbital phase.
		The orbital parameters are taken by considering the partial disruption condition.
		Panel (a) shows the effect of the orbital period on the flux density.
		Panels (b), (c), and (d) show the effects of the surface magnetic field
		and the clump size on the flux density.}
	\label{fig:fig3}
\end{figure}

As described in Section \ref{sec:model}, the orbits of the disrupted clumps will change due to
dynamical interactions. Here, for simplicity, when studying
their interaction with the pulsar wind, we only consider their first round of motion in the orbit.
We further assume that the orbit is similar to that of the original planet. Using the orbital
parameters constrained from the partial disruption condition, we can estimate the peak flux of
the FRB generated due to the interaction of a clump and the pulsar wind through Eq.~(\ref{wind->frb}).
As an example, we have applied our model to FRBs 180916 and 121102. 	
In our calculations, we take $ \Omega_{\rm A} = 0.1 $ sr, $ \gamma = 3\times10^{6} $, and $ \epsilon_{\rm r} = 10^{-2} $.
Figure~{\ref{fig:fig3}} shows the flux variation versus the orbital phase,
which is caused by the variation of the distance $ r $ between the
NS and the clump due to the large orbital eccentricity.
Panel (a) of Figure~{\ref{fig:fig3}} shows the effect of $ P_{\rm orb} $ on the flux density.
It is clear that the flux density is quite insensitive to $ P_{\rm orb} $ under the partial disruption condition.
Panels (b) -- (d) of Figure~{\ref{fig:fig3}} show the effects of $ B_{\star} $ and $ R_{\rm c} $ on $ S $ for
fixed $ P_{\rm orb} $ and $ P_{\star} $.
From these plots we can see that the effects of $ B_{\star} $ and $ R_{\rm c} $ are significant.
Note that for many parameter sets, $ S $ is larger than the detection threshold (0.3 Jy) during
a significant portion of the orbital phase.

	In a typical duty circle, the observed FRB number is of the order of a few. This indicates
	that usually only a few major fragments are generated during the passage of the periastron.
	For the clump-wind interaction mechanism, the required size of the small body that can produce FRBs is a few kilometers.
	It is quite typical for the fragments generated during a
	partial disruption. Simulations show that the tidal disruption of a planet by
	a compact star such as a WD can give birth to fragments ranging from a few
	kilometers to $\sim 100$ km \citep{Malamud2020a, Malamud2020b}.
	This is interesting to note that the number of observed bursts is related with fluence
	as $ N \propto F^{\alpha + 1}$,
	where $\alpha = -2.3\pm 0.3 $ for FRB 180916 \citep{Chime/Frb2020Natur}. So,
	there are many more low-fluence FRBs as compared with high-fluence ones.
	This is consistent with our NS-planet interaction model.
	In the partial disruption process, the number of smaller clumps is usually
	larger than that of bigger fragments \citep{Malamud2020a, Malamud2020b}.

\section{Periodicity and active window} \label{sec:periodicity-active}

\subsection{Periodicity} \label{subsec:periodicity}
Observations indicate that FRB 180916 seems to have a repeating period of
16.35 days \citep{Chime/Frb2020Natur}, while FRB 121102 may have a period of
157 days \citep{Rajwade2020MNRAS}. This suggests that the periods of repeating FRBs
may vary in a relatively wide range. In our model, the period is mainly determined
by the orbital motion of the planet. The observed periods thus exert some constraints
on the parameters of our NS-planet systems. Here we show that the planet-disruption
model can meet the observational requirements.

\begin{figure}[!h]
	\plotone{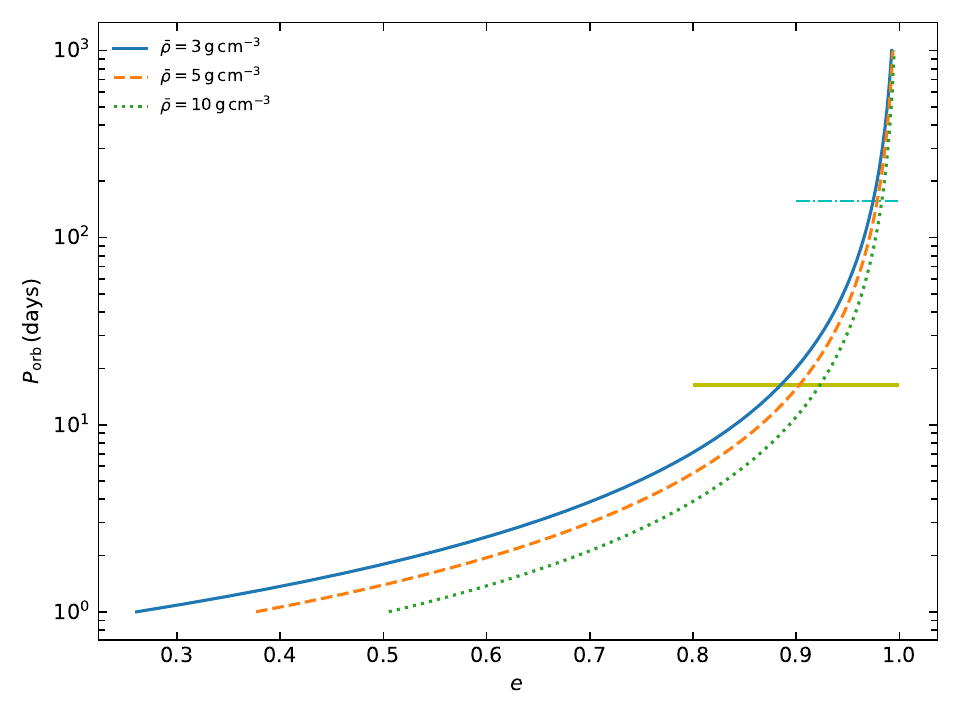}
	\caption{Orbital period as a function of the eccentricity under the partial tidal disruption condition
	 $ r_{\rm p} = 2\,r_{\rm td} $. The calculation is conducted for three different
		densities. The two horizontal short lines represent orbital periods of 16.35 days and 157
		days, respectively.}
	\label{fig:fig4}
\end{figure}

As mentioned in Section~\ref{sec:model}, we take $ r_{\rm p} = 2\,r_{\rm td} $ as
the typical case for the partial disruption condition. This naturally leads to a
relation of
\begin{eqnarray} \label{eq:td_condition}
	a\left(1 - e\right) = 2\,\left( \frac{6M_{\star}}{\pi\,\bar \rho}\right) ^{1/3}.
\end{eqnarray}
Combining Eq.~(\ref{eq:kepler_3}) and Eq.~(\ref{eq:td_condition}),
one can derive the relationship between various parameters of the NS-planet systems.
In Figure~\ref{fig:fig4}, we have plot the relation between the eccentricity and
orbital period for planets that satisfy the partial disruption condition.
The calculations are conducted for planets with a mean density
of $\bar \rho = 3 \rm \, g\, cm^{-3}$,
$5 \rm \, g\, cm^{-3}$, and $ 10 \rm \, g\, cm^{-3}$, respectively.
We can see that with an increase of the period, the eccentricity should also increase.
This is easy to understand. The key point is that the periastron distance
($ r_{\rm p} = 2\,r_{\rm td} $) is almost fixed by the mean density in our framework.
At the same time, to acquire a long orbital period, the semi-major axis should be
large enough. As a result, the eccentricity will have to be large. From Figure~\ref{fig:fig4},
we can see that to get a period of $\sim 1$ day, an eccentricity of $e \sim $ 0.3 --- 0.5
is enough. However, to achieve a period of $\sim 16$ days, $e \sim $ 0.9 will be
required, while for $P_{\rm orb} \geq 160$ days, $ e \geq 0.97$ is necessary.
In general, Figure~\ref{fig:fig4} demonstrates that partial disruption does can
happen periodically under proper conditions, and repeating FRBs with periods ranging
from $\sim 1$ to $\sim 1000$ days are possible.

\subsection{Active window} \label{subsec:active}
In the context of the Alfv\'{e}n wing mechanism, the active window of FRBs is determined by the distribution of
clumps in the orbit. The clumps originating from different parts of the planet have slightly different orbital parameters.
The semi-major axis of the clumps disrupted from a planet around a WD is given in \citet{Malamud2020a}.
Here, we applied it to our model as
\begin{equation}\label{a_prime}
	a^{\prime} = \left\{
	\begin{array}{lr}
		a \left( 1 + a\frac{2R}{d(d - R)}\right)^{-1},~~(\rm In~the~direction~of~NS)\\
		a \left( 1 - a\frac{2R}{d(d + R)}\right)^{-1},~~(\rm In~the~opposite~direction~of~NS)
	\end{array}
	\right.
\end{equation}
where $ a $ is still the planet's original semi-major axis, $ d $ is
the distance between the pulsar and planet at the moment of breakup, and $ R $ is the displacement
of the clump relative to the planet's mass center at the moment of breakup ($ R = 0 $
corresponds to the center of the planet).
In the opposite direction of the NS, there is a critical displacement $ R_{\rm crit} = d^{2}/(2a - d) $.
Particles with $ R < R_{\rm crit} $ are bound while particles with $ R > R_{\rm crit} $
are unbound to the planet \citep{Malamud2020a}.

The semi-major axes of disrupted clumps are different since their displacements ($ R $)
are different (see Eq. (\ref{a_prime})). Hence, their velocities and orbital periods are also different.
The orbital velocity can be calculated by $ \upsilon = r \omega $,
where $ \omega = ({2 \pi}/{P_{\rm orb}}){(1 + e \cos \theta)^2}/{(1 - e^2)^{3/2}} $ \citep{Sepinsky2007ApJ}.
Substituting $ r $ with Eq. (\ref{eq:r_orb}), we get
\begin{equation}\label{velocity}
	\upsilon^{2} = \frac{G(M_{\star} + m)}{a} \frac{(1 + e \cos\theta)^2}{(1 - e^2)}.
\end{equation}
This is the velocity of the planet at phase $ \theta $. We can further obtain the velocity of the
disrupted clumps by substituting $ a $ in Eq.~(\ref{velocity}) with $ a^{\prime} $ of Eq.~(\ref{a_prime}).

The active window of the wind interaction mechanism is determined by the difference of the
orbital periods of the clumps in the innermost and outermost orbits, which themselves can
be obtained by combining Eq. (\ref{a_prime}) and Eq. (\ref{eq:kepler_3}).
Here, we assume that the line of sight lies in the orbital plane.
In our calculations, for simplicity, we assume that the clumps are disrupted from the surface of the
planet (i.e. $ R = R_{\rm c} = (3m/4\pi\bar\rho)^{1/3} $) at the periastron
$ d = r_{\rm p} = 2 r_{\rm td}$.
As an example, we take a planet's parameters as $ P_{\rm orb} = 100 $ days, $ m = 10^{-6} \, M_{\odot} $, and
$\bar \rho = 5 \rm \, g\, cm^{-3}$ (correspondingly, $ e = 0.971 $).
The orbital velocity of such a planet at the periastron is $\rm 423.5 \,km\,s^{-1} $.
The velocity of the clumps in the outermost orbit is $\rm 389.5\,km\,s^{-1} $,
corresponding to an orbital period of $ P_{\rm orb}^{\rm out} = 128.5 $ days. For the clumps in the innermost orbit,
the velocity is $\rm 455.1\,km\,s^{-1} $, corresponding to an orbital period $ P_{\rm orb}^{\rm in} = 80.6 $ days.
We can see that the difference of their orbital periods is 47.9 days.
Below, we will consider the active windows of FRBs 180916 and 121102 in more detail.

\begin{figure}[!h]
	\plotone{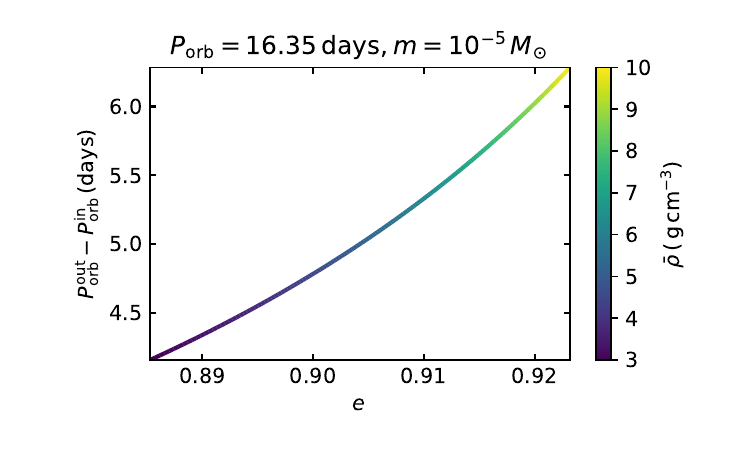}
	\caption{Difference of orbital periods for the clumps in the innermost and outermost orbits,
		plotted vs. the eccentricity for a planet in a 16.35 day orbit.}
	\label{fig:fig5}
\end{figure}

The repetition period of FRB 180916 is 16.35 days and the active window is about 5 days.
Figure~\ref{fig:fig5} shows the period difference for the clumps in the innermost
and outermost orbits as a function of the eccentricity and density. Under the partial disruption condition,
a planet with $ m = 10^{-5} \, M_{\odot} $ and $ P_{\rm orb} = 16.35 $ days can produce clumps with
period differences ranging from 4 to 6.25 days when the density ranges from
$\bar \rho = 3 \rm \, g\, cm^{-3}$ to $\bar \rho = 10 \rm \, g\, cm^{-3}$. In these cases, the velocity of
the planet is $\rm 429.7\,km\,s^{-1} $ at the periastron.
The velocity of the clumps in the outermost orbit is $ 406.4 \rm \,km\,s^{-1} $, corresponding to an orbital
period $ P_{\rm orb}^{\rm out} = 19 $ days. For the clumps in the innermost orbit, the velocity is
$\rm 451.9\,km\,s^{-1} $ and the orbital period is $ P_{\rm orb}^{\rm in} = 14 $ days.
The period difference is 5 days, which can satisfactorily meet the requirement of the observed active
window of FRB 180916.

\begin{figure}[!h]
	\plotone{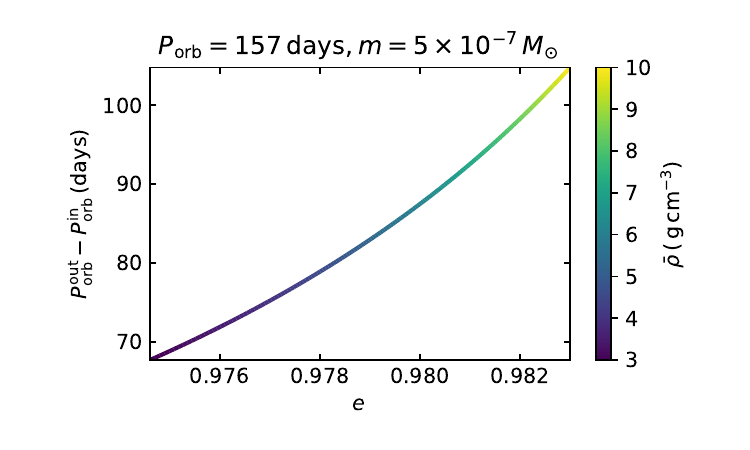}
	\caption{Difference of orbital periods for the clumps in the innermost and outermost orbits,
		plotted vs. the eccentricity for a planet in a 157 day orbit.}
	\label{fig:fig6}
\end{figure}

The repetition period of FRB 121102 is about 157 days, and the active window is  $\sim 87$ days.
Figure~\ref{fig:fig6} shows the period difference for the clumps in the innermost
and outermost orbit as a function of the eccentricity and density. Under the partial disruption condition,
a planet with $ m = 5 \times 10^{-7} \, M_{\odot} $ and $ P_{\rm orb} = 157 $ days can produce clumps with
period differences ranging from 65 days to about 105 days when the density
ranges from $\bar \rho = 3 \rm \, g\, cm^{-3}$ to $\bar \rho = 10 \rm \, g\, cm^{-3}$.
In these cases, the velocity of the planet is $\rm 437.57\,km\,s^{-1} $ at periastron.
The velocity of the clumps in the innermost orbit is $\rm 474.6\,km\,s^{-1} $, corresponding to an orbital
period $ P_{\rm orb}^{\rm in} = 123.06 $ days. For the clumps in the outermost orbit, the velocity
is $\rm 397.28\,km\,s^{-1} $ and the orbital period is $ P_{\rm orb}^{\rm out} = 209.78 $ days.
The period difference is 86.72 days. Again it can satisfactorily meet the requirement of the observed active
window of FRB 121102.

	Note that the gravitational perturbations from the planet itself still could influence the orbits of the clumps.
	Such perturbations can lead to changes in the inclination and eccentricity of the clump orbits so that the clumps will
	finally deviate from the line of sight. As a result, FRBs could be observed only from newly generated
    fragments, which maintains the periodicity and active window of the repeating FRBs. Below,
    we present more details on this issue.
	
	In a triple system where a test particle revolves around its host in a close inner orbit while
	a third object moves around in an outer orbit, the eccentricity of the test particle
	can be significantly altered by the outer object. This is called the Kozai-Lidov effect,
	which can change the orbit of the test particle \citep{Kozai1962,Lidov1962,Naoz2016ARAA}.
	In a normal Kozai-Lidov mechanism, the bigger planet's orbit is usually assumed to be circular
 	and the vertical angular momentum is conserved for the test particle.
	As a result, the eccentricity and inclination of the test particle's orbit vary periodically.
	However, when the planet's orbit is eccentric, the z-component
	of the inner and outer orbits' angular momentum is not conserved,
	which leads to very different behaviors of the test particle \citep{Lithwick2011ApJ, Li2014ApJ, Naoz2017AJ}.
	It was found that for a nearly coplanar (the inclination $ i \sim 0 $) and highly eccentric
	(for both inner and outer) configuration, the eccentricity of the test particle increases steadily,
	while the inclination $ i $ oscillates in a small range \citep{Li2014ApJ}.
	It was also found that, for a system with a tight-orbit configuration, the perturbation is strong and
	the orbit of the test particle can be altered on short timescales. 
	
	In our model, as mentioned above, the clumps coming from different parts of the
	planet move in slightly different orbits as compared with that of the planet. These orbits
	are approximately coplanar and close to each other. The surviving major portion of
	the planet can create perturbation.
	Unlike the case of \citet{Li2014ApJ}, our system breaks the secular approximation condition.
	Such a case has been discussed by \citet{Antonini2014ApJ}.
	They found that the inclination and eccentricity of the test object still
	could change in a short time. As a result, in our cases, the direction of the Alfv\'{e}n wing
    and the FRB emission cone will deviate from our line of sight in a short time
    (e.g., after one or two orbital periods) due to the inclination change.
    No FRBs would be observed from older fragments.

	To summarize, in our framework, the line of sight lies in the original orbital plane of the planet.
	The clumps generated during the partial disruption process near the periastron will
	pass through the observer's line of sight one by one during their first round of motion in their
    new orbits, producing FRBs detectable by the observer. After that, the gravitational
    perturbation from the planet will change the orbits of the fragments so that they will no
    longer produce visible FRBs later. In other words, only new clumps generated near the
    periastron will produce FRBs. In this way, the periodicity and active window of the repeating FRBs
    can be well maintained.

\section{Evaporation timescale for an object in an eccentric orbit}\label{sec:evaporation}

For a planet composed of ordinary matter orbiting around a pulsar, the evaporation timescale at a fixed distance is \citep{Kotera2016AA}
\begin{equation}\label{tev}
	t_{\rm ev} = 7.2 \times 10^{-12} {\rm yr} \left( \frac{R_{\rm c}}{\rm km}\right)^3 \left( \frac{\bar\rho}{\rm g\,cm^{-3}}\right)^2 \left( \frac{r}{R_{\odot}}\right)^2
	\frac{1}{L_{p,35}Q_{\rm abs}},
\end{equation}
where $ L_{\rm p} = 9.6\times10^{34} {\rm erg\,s^{-1}} I_{45} P_{\star,-3}^{-4} B_{\star, 8}^2 R_{\star ,6}^6 $
is the spin-down luminosity of the pulsar (here the convention $ Q_{x} = Q/10^{x} $ in cgs units is adopted),
and $ Q_{\rm abs} $ is the energy-absorption efficiency.
Usually, $ Q_{\rm abs} = 3/N_{\rm r}$ for large planets and $ Q_{\rm abs} = 12x/N_{\rm r}^2 + 2x^3N_{\rm r}^2/15$ for small objects.
Here, the refractive index is $ N_{\rm r} = \sqrt{\mu\sigma P_{\star}}$, with $ \mu\sigma = 10^6 \rm\,s^{-1}$
and the size ratio is defined as $ x = R_{\rm c}/cP_{\star} $.
Taking $\ P_{\star} = 1 \,\rm s $ and $ B_{\star} = 10^{12}\rm\, G $, the evaporation
timescale is $ t_{\rm ev} \sim 10^4 $ yr for a small object with a density of $\bar \rho = 3 \rm \, g\, cm^{-3}$
in a circular orbit of $ r \sim 10^{11}\rm\,cm $ \citep{Kotera2016AA}.
However, the situation is very different for an object in a highly eccentric orbit, because the distance
of the two objects varies in a very wide range. More importantly, for most of the time of each orbital cycle,
the planet is far away from the pulsar. In our model, the periastron distance is $\sim 10^{11} $ cm,
but the average separation in one orbital period is much larger than this value.
In the case of an elliptical orbit, the mean separation between the two objects is
$ \bar{r} = \frac{2}{P_{\rm orb}}\int_{0}^{P_{\rm orb}/2} r dt $ and can be calculated as
\begin{equation}\label{mean_distance}
	\bar{r} = \frac{a(1-e^2)^{5/2}}{\pi}\int_{0}^{\pi} \frac{1}{(1 + e \cos\theta)^3}d\theta.
\end{equation}

Substituting $ r $ in Eq. (\ref{tev}) with $\bar{r}$, we can estimate the evaporation timescale
of a planet in an elliptical orbit.
In our model, assuming $ m = 10^{-5}\,M_{\odot} $, $ P_{\rm orb} = 16.35\,\rm days $,
and $ \bar\rho = 3 \rm\,g \,cm^{-3} $, then the planet will be partially disrupt when $ e = 0.88 $.
In this case, the evaporation timescale can be derived as $ t_{\rm ev} = 2 \times 10^{7} $ yr.
If the mean density is taken as $ \bar\rho = 10 \rm\,g \,cm^{-3} $, then it will be partially disrupted
when $ e = 0.92 $ and the evaporation timescale correspondingly becomes $ t_{\rm ev} = 7 \times 10^{7} $ yr.
If we take the pulsar spin period as $ P_{\star} = 1 \rm\, s $, then we get $ t_{\rm ev} = 6.3 \times 10^{11} $ yr
for $ \bar\rho = 3 \rm\,g \,cm^{-3} $, and $ t_{\rm ev} = 2.2 \times 10^{12} $ yr for $ \bar\rho = 10 \rm\,g \,cm^{-3} $.
A planet with $ m = 10^{-5}\,M_{\odot} $, $ P_{\rm orb} = 157\,\rm days $ and $ \bar\rho = 3 \rm\,g \,cm^{-3} $
will be partially disrupted when $ e = 0.975 $, corresponding to an evaporation timescale $ t_{\rm ev} = 4.6 \times 10^{8} $ yr. If the mean density is taken as $ \bar\rho = 10 \rm\,g \,cm^{-3} $,
then it will be partially disrupted when $ e = 0.983 $, corresponding to an evaporation timescale
$ t_{\rm ev} = 1.5 \times 10^{9} $ yr. If we change the pulsar spin period to $ P_{\star} = 1 \rm\, s $,
then we get $ t_{\rm ev} = 1.6 \times 10^{13} $ yr for $ \bar\rho = 3 \rm\,g \,cm^{-3} $,
and $ t_{\rm ev} = 4.9 \times 10^{13} $ yr for $ \bar\rho = 10 \rm\,g \,cm^{-3} $.
From the above calculations, we can see that the evaporation timescale of a planet in our elliptical
orbit is generally very large. Therefore, the effect of evaporation is negligible in this framework.


\section {Formation of high-eccentricity planetary systems} \label{sec:formation of eccentric orbit}

In Section 3, we demonstrated that to account for the observed repeating FRB periods ranging from tens of days
to over one hundred days, a highly elliptical planet orbit with $ e \geq 0.9 $ is needed. It is a
natural question that whether such highly elliptical orbits are possible or not for planets. Here we
present some discussion on this issue.

Since the discovery of the first extrasolar planet around PSR 1257+12~\citep{Wolszczan1992Natur},
about 4700 exoplanets (as of 2021 April 27) have been discovered (see Extrasolar Planets Encyclopaedia -- EU
\footnote{\url{http://www.exoplanet.eu/}}; \citep{Schneider2011AA}). Among them, more than 10 objects
are pulsar planet candidates.  Although the eccentricities of these pulsar planet candidates are
generally very small,
high-eccentricity pulsar binaries have been discovered (see references in the databases ``Pulsars in globular clusters" \footnote{\url{http://www.naic.edu/~pfreire/GCpsr.html}}
and The ATNF pulsar catalog
\footnote{\url{https://www.atnf.csiro.au/research/pulsar/psrcat/}}; \citep{Manchester2005AJ}).
Additionally, a few planets with a large eccentricity orbiting around other types of stars have also
been detected (see the EU database). Good examples for these include HD 20782 b ($ e = 0.97\pm0.01 $),
HD 80606 b ($ e = 0.93366\pm0.00043 $), HD 7449 A b ($ e = 0.92\pm0.03 $),
and HD 4113 A b ($ e = 0.903\pm0.005 $).
The existence of these special planets indicates that the formation of high-eccentricity
planetary systems around compact objects should also be possible.
Planets with a large eccentricity could be formed around a NS through at least three channels.
First, a free-floating planet (FFP) can be captured by a NS when they are in a close encounter.
Second, exchange/snatch of a planet may happen between a NS and a nearby main-sequence planetary system.
Thirdly, the Kozai-Lidov effect in a multibody system may give birth to a high-eccentricity planet.
Below, we discuss these three processes briefly.

\begin{itemize}
	\item Formation from the capture of FFPs by NS:
	
	FFPs are common in space \citep{Smith2001MNRAS,Hurley2002ApJ,Sumi2011Natur,Elteren2019AA,Johnson_2020ApJ,Przemek2020ApJ}.
	They may be formed from various dynamical interactions (see Figure~1 in~\citet{Kremer2019ApJ}),
	such as ejection from dying multiple-star systems \citep{Veras2012MNRAS,Wang2015MNRAS,Elteren2019AA},
	planet-planet scattering \citep{Hong2018ApJ,Elteren2019AA},
	or the encounter of a star with other planetary systems \citep{Hurley2002ApJ}.
	In a cluster's full lifetime, about 10$\%$ --- 50$\%$ of primordial planetary systems
	experience various dynamical encounters and many planets become FFPs.
	About 30$\%$ --- 80$\%$ of them escape the cluster due to strong dynamical encounters and/or
	tidal interactions \citep{Kremer2019ApJ} and travel freely in space.
	The velocity of these FFPs is typically in the range of  0 --- 30 $\rm \,km\,s^{-1} $ \citep{Smith2001MNRAS,Hurley2002ApJ}.
	FFPs may be captured by other stars or planetary systems and form highly eccentric
	planetary systems \citep{Parker2012MNRAS,Wang2015MNRAS,Li2016ApJ,Goulinski2018MNRAS,Hands2019MNRAS,Elteren2019AA}.
	A simulation by \citet{Goulinski2018MNRAS} showed that more than 99.1$\%$ of the
	captured planets are in an orbit with $ e > 0.85 $, and the masses of FFPs do not affect the
	eccentricity significantly.

	\item Formation from NS exchange/snatch a planet:
	
	Pulsars can obtain a kick velocity when they are born in the supernova explosion.
	If a planet survives in supernova, the newborn high-speed pulsar and the surviving planet may form an
	eccentric planetary system by gravitational interaction. Additionally, when a pulsar moves with
	a kick velocity of  100 --- 400 $\rm \,km\,s^{-1} $ in space, it may pass by a planetary
	system. During this process,
	the pulsar can also exchange/snatch a planet from other planetary systems via
	gravitational perturbations. Planetary systems formed in this way may also be eccentric.
	
	\item Formation from the Kozai-Lidov effect in a multibody system:
	
	The Kozai-Lidov effect \citep{Kozai1962,Lidov1962,Naoz2016ARAA} can explain the dynamics of multibody systems
	in which one companion in an outer orbit can change (increase) the eccentricity of objects in inner orbits
	by gravitational perturbations. The timescale for forming a high-eccentricity system is determined by the
	initial parameters. If the central star of such a multibody system is a NS then a highly
	eccentric NS-planet system may form.

\end{itemize}

From the above descriptions, we can see that there are many routes to form high-eccentricity
planets around NSs. The requirement of $e \geq 0.9$ in our framework thus
in principle can be met in reality.

Here, we roughly calculate the population of highly eccentric planetary systems in the Milky Way.
It is estimated that there are 100 -- 400 billion stars in our Galaxy
(see the Universe Today\footnote{\url{https://www.universetoday.com/30296/how-many-planets-are-in-the-galaxy/}} and
NASA\footnote{\url{https://asd.gsfc.nasa.gov/blueshift/index.php/2015/07/22/how-many-stars-in-the-milky-way/}} websites).
A study based on the microlensing observations suggests that each star hosts
1.6 planets on average \citep{Cassan2012Natur}. Taking 200 billion as the rough number of stars, then there would be
about 320 billion planets in the Milky Way. Since about 10$\%$ --- 50$\%$ of primordial planetary systems
experience various dynamical encounters and produce FFPs as mentioned above \citep{Kremer2019ApJ}, it is
expected that there should be 20 -- 100 billion FFPs in the whole Galaxy.
More than $ 85\% $ of the stars in the Galactic disk are in a mass range of $ 0.1 M_{\odot} < M < 2 M_{\odot} $.
About $ 1\% $ of them are expected to experience at least one capture process during their
lifetime \citep{Goulinski2018MNRAS}. This allows us to estimate that there are 1.7 billion
captures and $ 99.1\% $ (1.68 billion) of them give birth to planets in a highly eccentric orbit with $ e > 0.85 $.
Currently, four highly eccentric ($e > 0.9$) planets have been confirmed among the observed 4700 planets,
corresponding to a fraction of $ 0.085\% $. Using this ratio as a reference, it can be estimated that
the number of highly eccentric ($ e > 0.9 $) planetary systems in our Galaxy is $\sim 170$ million.
From the above analysis, we can see that highly eccentric planetary systems are copious in the Milky Way.
However, it is not easy to detect them due to various observational biases. For these planets, the evaporation
again can be safely omitted since the timescale is usually much more than $ 10^{7} $ yr.


\section{Conclusions and discussion} \label{sec:conclusion}

In this study, we aimed to explain the periodic repeatability of FRBs by considering
a NS-planet interaction model.
In our framework, a planet moves around its host NS in a highly eccentric orbit.
The periastron of the planet satisfies a special condition 
$ r_{\rm td} \leq r_{\rm p} \leq 2\,r_{\rm td} $, so that the crust of the planet
will be partially disrupted every time it pass through the periastron. Fragments of the
size of a few kilometers are produced in the process.
During the process, the fragments interact
with the pulsar wind via the Alfv\'{e}n wing mechanism to give birth to FRBs.
The periods of repeating FRBs correspond to the orbit periods of
the planets. To account for the observed period of $\sim 10$ --- 100 days, an orbital
eccentricity larger than $\sim 0.9$ is generally required.
It is shown that the basic features of the two well-known repeating sources, FRBs 121102
and 180916, can be satisfactorily interpreted by the model.

It is interesting to note that the interaction of small bodies with NSs has already been
studied to interpret repeating FRBs, but generally in s very different framework.
For example,
\citet{Dai2016} explained repeating FRBs as due to the multiple collisions that happen
when a NS travels through an asteroid belt.
\citet{Decoene2021AA} even suggested a three-component scenario which
involves a NS, an asteroid belt around it, and a third outer companion.
In their model, the outer companion can be a black hole, a NS, a WD or a main-sequence star.
While our model is in principle different, we would like to point out that some
ingredients in the above models may also play a role in our model.
For example, when the fragments finally arrive at the NS and collide with it, FRBs may be produced via
the NS-asteroid collision mechanism \citep{Geng2015,Dai2016}.
Yet, the time needed for the clumps to fall into the NS is highly uncertain and still needs to be further studied.
Note that the disruption distance of rocky planets is
$ \sim 10^{11}\rm cm $ \citep{Mottez2013AA1,Mottez2013AA2}.
At this distance, the evaporation takes a time of only $\sim 10^{4} $ yr \citep{Kotera2016AA}.
However, the ellipticity of the orbit can prolong the evaporation timescale
by several orders of magnitude, to $\geq 10^{7} $ yr. Therefore, the evaporation does
not affect our model significantly.

\section{Acknowledgments}

We would like to thank the anonymous referee for helpful suggestions that led to
significant improvement of our study.
This work is supported by the special research assistance project of the Chinese
Academy of Sciences (CAS), by the National Natural Science Foundation of China (grant Nos. 12041306, 12041304,
11873030, U1938201, U1838113, U2031209, 11903019, 11833003, 12033001, 12103055),
by the National Key R\&D Program of China (grant No. 2021YFA0718500),
by the National SKA Program of China (grant No. 2020SKA0120300),
by the CAS ``Light of West China'' Program (grant No. 2018-XBQNXZ-B-025),
by the science research grants from the China Manned Space Project with No. CMS-CSST-2021-B11,
and by the Operation, Maintenance and Upgrading Fund for Astronomical Telescopes and Facility Instruments,
budgeted from the Ministry of Finance of China (MOF) and administrated by the CAS.
This work is also partially supported by the Strategic Priority Research Program
of the CAS under grant No. XDA15360300.
\software{SciPy \citep{Virtanen2020}, Matplotlib \citep{Hunter2007}, NumPy \citep{Walt2011}}.


\nocite{*}
\bibliographystyle{aasjournal}
\bibliography{reference}

\end{CJK*}
\end{document}